# Descriptor for slip-induced crack-blunting in refractory ceramics

Davide G. Sangiovanni,[1*] Antoine Krych,[2] Matous Mrovec,[2] Janella Salamania,[1] Magnus Odén,[1] Ferenc Tasnádi,[1] Igor A. Abrikosov[1]

[1] Department of Physics, Chemistry, and Biology (IFM), Linköping University, SE 58183 Linköping, Sweden

[2] The Interdisciplinary Centre for Advanced Materials Simulation (ICAMS), Ruhr-Universität Bochum, D-44780 Bochum, Germany

Understanding the competition between brittleness and plasticity in refractory ceramics is of importance for aiding design of hard materials with enhanced fracture resistance. Inspired by experimental observations of crack shielding due to dislocation activity in TiN ceramics [Kumar et al., International Journal of Plasticity **27** (2011) 739], we carry out comprehensive atomistic investigations to identify mechanisms responsible for brittleness and slip-induced plasticity in Ti-N systems. First, we validate a semi-empirical interatomic potential against density-functional theory results of Griffith and Rice stress intensities for cleavage ($K_{Ic}$) and dislocation emission ($K_{Ie}$) as well as *ab initio* molecular dynamics mechanical-testing simulations of pristine and defective TiN lattices at temperatures between 300 and 1200 K. The calculated $K_{Ic}$ and $K_{Ie}$ values indicate intrinsic brittleness, as $K_{Ic} \ll K_{Ie}$. However, $K_I$-controlled molecular statics simulations − which reliably forecast macroscale mechanical properties through nanoscale modelling − reveal that slip-plasticity can be promoted by a reduced sharpness of the crack and/or the presence of anion vacancies. Classical molecular dynamics simulations of notched Ti-N supercell models subject to tension provide a qualitative understanding of the competition between brittleness and plasticity at finite temperatures. Although crack growth occurs in most cases, a sufficiently rapid accumulation of shear stress at the notch tip may postpone or prevent fracture via nucleation and emission of dislocations. Furthermore, we show that the probability to observe slip-induced plasticity leading to crack-blunting in flawed Ti-N lattices correlates with the *ideal* tensile/shear strength ratio ($I_{plasticity}^{slip}$) of pristine Ti-N crystals. We propose that the $I_{plasticity}^{slip}$ descriptor should be considered for ranking the ability of ceramics to blunt cracks via dislocation-mediated plasticity at finite temperatures.

*Corresponding author: davide.sangiovanni@liu.se

# 1. Introduction

The functionality of ceramics at temperatures ($\gtrsim$1000 K) of relevance for industrial, engineering, automotive, and nuclear energy applications is generally limited by the inherent tendency to crack (brittleness) of these materials [1] and/or by thermally-induced reductions in hardness [2] and strength [3]. These problems are of significant technological, economic, and practical concern [4], especially in safety-critical applications [5, 6]. Rational design of ceramics that combine strength *and* toughness (enhanced fracture resistance) requires identification and understanding of plastic deformation at different length scales and temperatures. This is a very complex task because (*i*) plasticity typically oppositely affects strength and toughness, (*ii*) *in situ* experimental characterization of mechanical properties is often unfeasible, (*iii*) *ab initio* simulations over length-scales sufficient to accommodate extended defects – which control the mechanical behavior of real ceramics – would be overly time consuming. However, molecular dynamics simulations based on force-field models carefully trained upon *ab initio* and experimental data offer possibilities to address this challenge.

Recent *ab initio* studies [7-11] aimed at identifying electronic-structure origins of toughness and hardness – often correlated to strength in solids [12, 13] – in transition-metal carbonitrides; a class of refractory ceramics widely used in high-temperature machining and automotive applications. It has been shown that valence-electron concentration tuning can be employed to control the inherent hardness [8-10], strength [14], ductility and toughness [7, 15] of rocksalt-structure (B1) single-crystal defect-free phases. Nonetheless, extended structural defects as dislocations, grain boundaries, and nanosized cracks are inevitably present in industrially-manufactured bulk or protective thin-film coating ceramics. During use, these defects may act as weak links, i.e., sites for crack nucleation and growth. Accordingly, the strength and toughness of defect-free crystals [16, 17] are *ideal* upper-limits of properties attainable in real ceramics. The role played by extended defects needs to be addressed separately and requires substantial efforts. For this reason, it is desirable to identify properties inherent to single-crystals that can readily reflect trends in actual mechanical responses at finite temperatures and over length scales of practical relevance.

Griffith ($^G$) and Rice ($^R$) theories, based on linear elasticity, provide useful predictions of the stress intensity factors (K) required for unstable crack growth ($K_{Ic}^G$) or dislocation emission during mode-I ($K_{Ie}^R$) and mode-II ($K_{IIe}^R$) loading [18, 19]. Definitions of K, $K_{Ic}$, $K_{Ie}$, and $K_{IIe}$ can be found in the Supplemental Material (**SM**) [20]. A material with given crystallographic orientation is indicated to be intrinsically brittle during mode-I loading if $K_{Ic}^G < K_{Ie}^R$. The criteria for cleavage and emission are based on the anisotropic elastic response of the material, surface energies $E_s$, and unstable stacking fault energies $\gamma_{usf}$. However, the effective material resistance to crack growth or ability to emit dislocations includes effects that are often not well encompassed by elementary properties such as surface and stacking fault energies [18, 21, 22]. Moreover, $E_s$ and $\gamma_{usf}$ are quantities difficult to compute at finite temperatures. Thus, direct molecular statics and molecular dynamics simulations are required to achieve comprehensive understanding of material mechanical responses under load.

Recent experimental investigations revealed surprising resistance to fracture mediated by dislocation nucleation and/or motion in (typically brittle) TiN [23]. Inspired by that study, here we employ density-functional theory (DFT), molecular statics (MS), *ab initio* and classical molecular dynamics (AIMD and CMD) to gain qualitative understanding of the role played by large structural defects (cracks and notches) on the mechanical properties and behavior of TiN and TiN$_x$ refractory ceramics. Specifically, we focus on identifying atomistic mechanisms that control the material ability to blunt cracks due to dislocation emission (slip-induced plasticity) in defective supercell models. Moreover, we introduce the $I_{plasticity}^{slip}$ descriptor (based on properties of defect-free single crystals) that correlates with the likelihood of occurrence of slip-induced plasticity at finite temperatures. Results presented for hard Ti-N crystals [24, 25] may serve for interpretation of the mechanical behavior of defective B1 carbonitrides in general [15, 26-30].

Core findings of brittleness vs plasticity mechanisms and correlation between $I_{plasticity}^{slip}$ and probability to activate slip-plasticity leading to crack blunting at finite temperatures are discussed in **Section 3.2**. In **Section 3.1**, we validate the semi-empirical potential against *ab initio* and classical simulation results obtained for single crystals and *small* notched supercells.

## 2. Computational details and methods

In this work, we investigate mechanical and extended-defect properties of B1 TiN and of a vacancy-ordered variant of cubic $TiN_x$ (see **Sec. 2.2.A** for details). DFT calculations and AIMD simulations are carried out using VASP [31], implemented with the projector augmented wave method [32]. In static DFT calculations, we employ both the local-density approximation (LDA) and the generalized-gradient approximation (GGA) of Perdew-Burke-Ernzerhof (PBE) [33] for description of electronic exchange and correlation energies and one-electron potential. We use the PBE functional in all AIMD simulations. The density of k-point grids and planewave cutoff energies differ depending on the property under investigation (see following sections). MS calculations and CMD simulations are performed using LAMMPS [34] and the modified embedded atom method (MEAM) [35]. The Ti-Ti, N-N, and Ti-N interactions are parameterized as described in Ref. [36]. The interaction cutoff is set to 7.1 Å, which is suitable to model tensile deformation. Both AIMD and CMD simulations integrate the equations of motion on time steps of 1 fs. Snapshots and videos of calculations and simulations are generated with VMD [37].

### 2.1. Static calculations

**2.1.A. Evaluation of $K_{Ic}^G$, $K_{Ie}^R$, and $K_{IIe}^R$ based on Griffith and Rice criteria.** The competition between brittle cleavage and dislocation emission for a given crystal orientation and crack-front direction during mode-I loading (opening) can be predicted at temperature T = 0 K from anisotropic linear elasticity theory using the criteria of Griffith and Rice. Rice criterion can also be used to predict the stress intensity $K_{IIe}^R$ required to nucleate and move a dislocation along the crack plane during mode-II loading (in-plane shear) of a pre-cracked solid. The theory has been comprehensively reviewed in Refs. [18, 19]. However, chapters from Ting's book [38] facilitate implementation.

Briefly, calculations of $K_{Ic}^G$, $K_{Ie}^R$, and $K_{IIe}^R$ require knowledge of the material elastic tensor $C_{ij}$, surface energy $E_s$ of the crack plane, and unstable stacking fault energy $\gamma_{usf}$ along a slip plane inclined by an angle $\theta$ to the fracture surface with Burgers vector orthogonal to the crack-front direction:

$$K_{Ic}^G = [\, 2\, E_s\, \Lambda_{22}^{-1}\, ]^{½}, \qquad (1)$$

$$K_{Ie}^R = [\, \gamma_{usf}\, \Lambda_{11}^{\theta\,-1}\, ]^{½} / F_{12}(\theta), \qquad (2)$$

$$K_{IIe}^R = [\, \gamma_{usf}\, \Lambda_{11}^{-1}\, ]^{½}. \qquad (3)$$

In the case of a rocksalt-structure lattice with (001) fracture plane and [010] crack-front direction, the slip systems with highest Schmid factor $m_{Schmid}$ (see definitions and **Eqs. S1–S3** in **SM** [20]) during mode-I loading are $(\bar{1}01)[101]$ and $(101)[10\bar{1}]$. Thus, the most likely emission planes are inclined $\theta = \pm 45°$ to the fracture plane. In **Eqs. (1,2)**, $\Lambda_{22}^{-1}$ and $\Lambda_{11}^{\theta\,-1}$ are elements of the inverse of $\overline{\overline{\Lambda}}$ and $\overline{\overline{\Lambda^\theta}}$ matrixes. $\overline{\overline{\Lambda}}$ is the Stroh energy tensor, which can be calculated solving an eigenvalue equation and knowing the material elastic tensor $\overline{\overline{C}}$ [19]. $\overline{\overline{\Lambda^\theta}}$ is the Stroh energy tensor in a coordinate system rotated about the crack-front axis by an angle $\theta$. $F_{12}(\theta)$ is a geometrical factor [19]. For evaluation of $K_{IIe}^R$, we consider a B1 lattice with crack plane on (110), crack front parallel to [001], and emission on $(110)[1\bar{1}0]$.

The surface energy (unrelaxed and relaxed) and generalized stacking fault energy surface are obtained by DFT and MS following standard 2D-periodic supercell calculation procedures (see, for example, Ref. [39] for $E_s$ and [40] for $\gamma_{usf}$). In DFT, we converge $E_s$ and $\gamma_{usf}$ values with respect to k-point grid thicknesses and cutoff energies. The elastic constants of TiN obtained by MEAM and DFT+GGA are taken from Refs. [36] and [7]. Here, we also compute the elastic properties using the LDA approximation. As shown in **Sec. 3.1**, GGA and LDA values can be considered as lower and upper bounds for $K_{Ic}^G$, $K_{Ie}^R$, and $K_{IIe}^R$. **Table I** summarizes results of $C_{ij}$, $E_s$, $\gamma_{usf}$, $K_{Ic}^G$, $K_{Ie}^R$, and $K_{IIe}^R$.

**2.1.B. K-controlled MS simulations.** The effective resistance to brittle cleavage and dislocation emission in TiN and TiN$_x$ is assessed via K-controlled ($^{KC}$) MS calculations of cracked-plate models with sizes up to $\approx 1.2 \times 10^6$ atoms. K-controlled simulations portray the macroscale mechanical response of a crack subject to remote stresses from an atomistic perspective. They provide reliable fracture toughness $K_{Ic}^{KC}$ values and allow studying the effect of crack sharpness and geometry on the competition between brittle cleavage and emission.

Theory and methods are detailed in [18, 19]. In short, we employ square TiN and TiN$_x$ plates with: (*i*) (001) crack-surface, [010] crack-front direction for mode-I and (*ii*) (110) crack-surface, [001] crack-front direction for mode-II. The supercell is periodic along the crack-front direction. The thickness is ≈1.7 nm. Atoms in the frame region centered at the crack tip are sequentially displaced by incrementing the crack-tip stress intensity $K_I$ or $K_{II}$ at steps of 0.02 MPa √m. Calculation of the displacements requires computing the complex roots with positive imaginary part of a characteristic 4$^{th}$-order equation, where the coefficients depend on the elastic compliance tensor rotated according to the orientation of the defective lattice. All, except frame atoms, are relaxed via conjugate-gradient energy minimization at each K increment. The tolerances are set to $10^{-14}$ for relative energy changes and $10^{-14}$ eV/Å for forces. The interactions among atoms on opposite sides of the crack plane are screened over ≈1 nm. For mode-II loading we test only atomically-sharp cracks. For mode-I, we perform simulations with crack planes of 0, 1, 2, 3, and 5 atomic-layer heights. No atoms are removed for zero-height cracks. Half-rows of atoms are deleted in the other cases. The plates are constructed with equal number of atomic layers above and below the crack. We investigate plate-area effects on computed quantities (see **Sec. 3.1**).

**2.1.C. MS calculations of edge-dislocation core properties.** As will be discussed in the results **Section 3**, K-controlled MS and tensile-strain CMD simulations reveal nucleation and emission of $\{110\}\langle1\bar{1}0\rangle$ edge dislocations from the crack tip of flawed TiN and TiN$_x$ lattices. These observations motivate MS calculations of the energy required to nucleate (core line-energy) and stress necessary to move (Peierls stress [41]) dislocations in TiN vs TiN$_x$ systems. Proper evaluation of these properties would require corrections that compensate for spurious dislocation-dislocation elastic interactions (see, e.g., Refs. [42, 43]). Here, we qualitatively compare the properties of edge dislocations in TiN and TiN$_x$ neglecting elastic interactions. Estimates of line-energies and Peierls stresses aid interpretation of results to be presented in **Sec. 3**.

The initial dislocation-core structures are created using Atomsk [44]. The size of supercells varies between $10^5$ and $10^6$ atoms. The supercells contain two mirrored dislocation dipoles aligned perpendicular to the glide plane (see, e.g., figure 1 in [43]). The relationships between Cartesian and crystallographic axes are: x// Burgers vector direction [1$\bar{1}$0], y // dislocation-line direction [001], and z // slip-plane normal direction [110]. Conjugate gradient energy minimization is used to relax the atomic positions and the supercell aspect ratio while imposing orthogonality of the simulation box.

For representation of the dislocation core, the Nye tensor was calculated following Hartley et al. [45] with a cutoff of 3 Å and an exclusion angle of 15 degrees. The Nye tensor is the gradient of the inverse of the transpose of the lattice distortion. The lattice distortion is calculated as difference between the atomic positions in a lattice containing a dislocation and in the perfect crystal. We qualitatively compare the dislocation-core line-energies of TiN and TiN$_x$. The line energies are obtained by subtracting the total energy of a defective supercell from the energy of a dislocation-free supercell of the same stoichiometry. Converged energy per unit length is achieved by sequentially increasing the lateral supercell size – from ≈5 to ≈50 nm along the x direction – to enlarge the dislocation-free regions. For Peierls stress estimations, simple-shear strain of the plane normal to the dislocation line is incremented at steps of 0.02% until dislocation motion is observed.

**2.2. Ab initio and classical molecular dynamics**

**2.2.A. Determination of single-crystal properties.** AIMD simulations are based on Γ-point sampling of the reciprocal space and 300 eV cutoff energies. AIMD tensile and shear deformation of single-crystal supercells is modelled at temperatures of 300, 600, 900, and 1200 K using 576-atom supercells, as detailed in previous studies [16, 17]. Supercell elongation parallel to ⟨001⟩, ⟨110⟩, and ⟨111⟩ crystal axes – i.e., orthogonal to the easiest cleavage planes [46] – is sequentially increased up to fracture. {110}⟨1$\bar{1}$0⟩ simple-shear deformation is incremented until occurrence of lattice slip.

Among slip systems known to operate in B1-structure ceramics, the $\{110\}\langle 1\bar{1}0\rangle$ is experimentally-identified as most active in TiN [47, 48] and indicated by DFT calculations to exhibit the lowest Peierls stress [49].

We use CMD simulations to compute theoretical strengths, investigate fracture and slip-induced plasticity in B1 TiN and a vacancy-ordered TiN$_x$ (x=93.75) variant as a function of temperature. Vacancy ordering allows to maintain the same point-group symmetry in supercells with different orientations. During thin-film growth, vacancy formation or incorporation occurs at random positions [50]. Thermodynamics, however, can produce short- and long-range vacancy-ordering. Ordered TiN$_x$ variants have been predicted by DFT to be stable at low temperature [51] and experimentally-reported for TiN$_{0.82}$ [52-54] and TiN$_{0.45-0.61}$ [52, 55] single-crystals. Although the TiN$_{0.9375}$ structure employed herein has not been previously observed, it serves (in addition to TiN) to demonstrate correlation between mechanical properties of single-crystal and defective lattices.

Single-crystal supercells are initially equilibrated via NPT sampling. Then, tensile and shear strain is incremented by 2% every 3 ps, as in AIMD simulations [16, 17]. The maximum stresses withstood during AIMD and CMD mechanical-testing correspond to *ideal* tensile $\sigma_T$ or shear $\gamma_S$ strengths. We carry out three sets of CMD simulations of tensile and shear strain at each investigated temperature for improving the accuracy of $\sigma_T$ and $\gamma_S$ values.

**2.2.B. Modeling fracture of *small* flawed supercells.** The resistance to brittle fracture of stoichiometric B1 TiN crystals is initially assessed using both AIMD and CMD tensile-testing of flawed lattice models. The simulations are done at 300, 600, 900, and 1200 K. AIMD supercells are comprised of 1100 atoms (**Fig. S1a** of the **SM** [20]). They are formed of 546 N, 546 Ti, and 8 He atoms (helium not shown in **Fig. S1a**). The He atoms – kept at fixed positions in the middle of the two cavities – prevent crack healing during equilibration. A vacuum region of 11 Å separates supercell replicas along the *x* // [100] direction. In CMD, crack-healing is prevented by screening the interactions among atoms on different sides of the cavities. Both AIMD and CMD supercells are periodic along *y* (crack line direction) and *z* (tensile elongation direction), **Fig. S1**. CMD supercells are ×3 thicker along *y* (**Fig. S1b**). The structural parameters of notched models are set equal to those determined for single-crystals (**Sec. 2.2.A**). Mode-I loading is mimicked by incrementing the [001]-length of the simulation box up to 40%, with 2% strain-steps. At each strain, the structures are NVT-equilibrated at a target temperature for 1.6 ps in AIMD and 10 ps in CMD.

**2.2.C. CMD statistics of mechanical behavior of notched crystal models.** The reliability of the MEAM interactions [36] used in this work is confirmed by comparison with results of AIMD simulations and DFT calculations (**Sec. 3.1**). Thus, computationally-efficient CMD is used to collect statistics on the mechanical behavior (brittle vs plastic) of notched B1 TiN and TiN$_x$ supercells subject to tensile deformation. All supercells have lateral (*b*) and vertical (*h*) sizes of $12.4 \times 12.8$ nm$^2$, notches of equal shape, but different sizes of a void region (V⊢⊣ = 0, 0.35, 0.40, 0.45, 0.50, 0.55, and 0.60*b*, **Fig. 1**). The void width is varied to produce different stress conditions at the notch tip during tensile strain. Notched supercells of 2.6 nm thickness are used in CMD simulations at 300, 600, 900, and 1200 K. For simulations at room temperature, we also test notched supercells with thicknesses of 0.9, 5.9, 8.5, and 17.0 nm (up to 250000 atoms). At least 70 statistically-independent simulations are run for a given supercell thickness at each temperature. Following equilibration, [001]-elongation of the simulation box is incremented at a rate of 1% every 10 ps. Atomic stresses at timestep *t* are time-averaged (smoothed) over ±1 ps.

## 3. Results and discussion
### 3.1. Validation of semi-empirical potential for intrinsic material properties
**3.1.A. Theoretical $K_{Ic}^G$, $K_{Ie}^R$, $K_{IIe}^R$, and K-controlled MS simulations.** According to Griffith and Rice criteria, a crystal lattice with given crack surface and crack front direction is prone to brittle cleavage during mode-I loading if $K_{Ic}^G < K_{Ie}^R$. Nucleation and emission of dislocations delays or prevents cracking otherwise. $K_{Ic}^G$ and $K_{Ie}^R$ are computed using **Eqs. 1,2** for TiN and TiN$_x$. For mode-I loading, we consider cracks on (001) with [010] front direction. We also derive $K_{IIe}^R$ for in-plane shear loading (crack on (110) plane) using **Eq. 3**. MS calculations yield the elastic constants, surface energies $E_s$, fracture energy $E_{s,\,unrel}$, and the unstable stacking fault energy $\gamma_{usf}$ of the slip system $\{110\}\langle 1\bar{1}0\rangle$, which has the highest Schmid factor during [001] tension. The $K_{Ic}^G$, $K_{Ie}^R$, and $K_{IIe}^R$ of TiN are also computed via DFT+PBE and DFT+LDA. The low accuracy of semi-local functionals (as PBE or GGA in general) for surface energy predictions has been discussed in various studies [56-58]. PBE typically underestimates surface energies, while LDA (although accidentally) is often more accurate [57-61]. On the other hand, the LDA overestimates the elastic constants and consequently also $\Lambda_{jj}^{-1}$ in **Eqs. 1–3**. Thus, PBE and LDA values define reasonable lower and upper bounds to $K_{Ic}^G$, $K_{Ie}^R$, and $K_{IIe}^R$. **Table I** shows that MS results are within or slightly off PBE and LDA ranges of $C_{ij}$, $E_{s,unrel}$, and $\gamma_{usf}$. More importantly, our classical potential predicts $K_{Ic}^G$, $K_{Ie}^R$, $K_{Ic}^G/K_{Ie}^R$ ratios, and $K_{IIe}^R$ with accuracy comparable to DFT (**Table I**). As expected, Griffith and Rice criteria indicate that both TiN(001)[010] and TiN$_x$(001)[010] are brittle, as $K_{Ie}^R \approx 2\,K_{Ic}^G$. A comparison of the ratios $K_{Ic}^G/K_{Ie}^R$ indicate that dislocation nucleation and motion is (in principle) more facile in the understoichiometric compound (**Table I**).

$K_I$-controlled MS simulations confirm the brittle behavior of TiN ceramics, as the crack propagates along the (001) surface at $K_{Ic}^{KC} = 1.68\pm0.01$ MPa $\sqrt{m}$ (pure screening) and at $K_{Ic}^{KC} = 1.78\pm0.01$ MPa $\sqrt{m}$ (atomically-sharp crack, see **Fig. 2a,b**). The $K_{Ic}^{KC}$ values are 8% and 14% greater than $K_{Ic}^G$ (=1.56 MPa $\sqrt{m}$, **Table II**). TiN$_x$ with atomically-sharp crack exhibits a $K_{Ic}^{KC}$ (1.82±0.01 MPa $\sqrt{m}$) that is nearly equivalent to that of TiN with the same crack shape and ≈20% greater than Griffith's value (**Table II**). The targeted $K_{Ic}^{KC}$ accuracy is reached with plates of area ≈4500 nm$^2$. Invariance of $K_{Ic}^{KC}$ values is verified using plate sizes up to ≈7300 nm$^2$ (≈1.2 × 10$^6$ atoms, see **Fig. S2**). We also control that the nonlinear material response is mainly confined near the crack tip (**Fig. S3**). The map in **Fig. S3** is obtained by the difference of relaxed and linear-elastic atomic displacements for $K_I$ = 1.40 MPa $\sqrt{m}$, which is close to critical stress-intensity values.

In contrast to the brittle behavior of plates with atomically-sharp cracks, TiN(001)[010] plates (area ≈7300 nm$^2$) with 2, 3, and 5 atomic-layer-thick cracks evidence nucleation of a $\{110\}\langle 1\bar{1}0\rangle$ edge dislocation at the crack tip. For the 2-atom-thick case, a dislocation forms at $K_I^{KC}$ = 1.84 MPa

√m but remains confined at the crack tip (**Fig. S4**). Presence of the dislocation delays fracture up to the critical value $K_{Ic}^{KC}$ = 2.20 MPa √m (**Fig. S4**). Our $K_{Ic}^{KC}$ results are consistent with experimental fracture toughness $K_{Ic}$ assessed by TiN microcantilever bending (**Table II**). In the plate with a 3-atom thick crack, a dislocation nucleates at $K_I^{KC}$ = 1.96 MPa √m (**Fig. 2c**) and moves toward the material interior for $K_I^{KC}$ = 2.04 MPa √m (**Fig. 2d**), indicating improved plastic behavior. A 5-atom thick crack results in nucleation *and* emission of a dislocation from the tip at $K_I^{KC}$ = 2.04 MPa √m (**Table II**).

The results of $K_I$-controlled MS simulations show that the sharpness of the crack qualitatively influences the mechanical behavior of TiN ceramics. Crack propagation is observed for crack-layers of 0 and 1-atom height. Nucleation, but not emission of dislocations is observed for crack-layers of 2-atom height. Dislocation nucleation *and* emission is observed for cracks of 3 and 5 atomic height. It is also important to note that the material mechanical behavior appears to be controlled by the crack-plane thickness alone, that is, qualitatively unaffected by the plate area. This is verified for plates with areas as small as ≈72 nm$^2$ (**Fig. S2**). For Ti-N systems, supercells of relatively small size correctly capture brittleness or plasticity mechanisms observed in large cells (≈7300 nm$^2$).

The $K_{IIe}^{KC}$ stress intensities for dislocation emission are recorded during $K_{II}$-controlled shearing of TiN(110)[001] plates with an atomically-sharp crack and area ≈7300 nm$^2$. MS simulations, consistent with Rice criterion, show that $K_{IIe}^{KC,\,TiN} > K_{IIe}^{KC,\,TiNx}$ (**Table II**). $K_{IIe}^{KC}$ results, together with $K_{Ic}^G/K_{Ie}^R$ values in **Table I**, suggest that the presence of anion vacancies assists nucleation and emission of dislocations from the notch tip upon loading. The predictions of MS simulations qualitatively agree with the observations of CMD simulations (**Sec. 3.2**) showing more frequent slip-induced crack blunting in TiN$_x$ than TiN.

**3.1.B. Properties of bulk single-crystals and $I_{plasticity}^{slip}$ descriptor.** The mechanical response to tensile elongation of stoichiometric single-crystal B1 TiN at room temperature has been recently investigated by AIMD simulations [16]. The results showed that the material undergoes brittle cleavage at the yield point, irrespective of the elongation direction [16]. In this work, we perform additional AIMD simulations to investigate the effect of temperature on the mechanical response of B1 TiN. AIMD stress/strain curves evaluated for tensile strain along ⟨001⟩, ⟨110⟩, and ⟨111⟩ are illustrated in **Fig. 3a,c,e,g**. The sudden drops in $\sigma_{zz}$ tensile stresses beyond yield points indicate occurrence of sudden bond-snapping leading to brittle cleavage. Exceptions are ⟨001⟩ elongation tests at 600 and 1200 K, for which fracture is preceded by gradual bond fraying. Nevertheless, the fracture mechanisms observed at different temperatures are similar to what already reported in Ref. [16].

The results of CMD simulations (**Fig. 3b,d,f,h**) are consistent with those of AIMD and suggest that single-crystal B1 TiN is an inherently brittle material. Considering that the MEAM parameters used in CMD had not been optimized for mechanical responses beyond the elastic limit [36], the agreement between CMD and AIMD results is remarkably good (**Fig. 3 and Table III**). CMD tensile-testing simulations are also used to determine the stress/strain relationships for a vacancy-ordered variant of understochiometric B1 $TiN_x$ (x=0.9375). The intent of identifying correlations between single-crystal and defective-crystal mechanical responses benefits from an accurate characterization of the properties of different materials. Testing of $TiN_x$ serves this purpose, because the MEAM potential used in this work accurately describes both stoichiometric and N-deficient TiN systems [36]. In addition, vacancy-ordering allows us to reproduce the same crystal symmetry in supercells with different crystallographic orientations relatively to the Cartesian axes.

Analogously to the stoichiometric phase, CMD simulations show that pristine $TiN_x$ single crystals cleave in all tests. CMD observations would suggest that both TiN and $TiN_x$ are equally brittle. Nevertheless, the actual material's resistance to fracture is affected by its ability to nucleate and emit dislocations from crack tips. This ability can be assessed using the ratio of surface energies ($E_{surf}$) and unstable stacking fault energies ($\gamma_{usf}$) [62-64], which are concepts founded in Griffith and Rice pioneering works [65, 66]. $E_{surf}$ is an estimate of the energy necessary to propagate a crack, whereas $\gamma_{usf}$ indicates the energy necessary to nucleate a dislocation. Here we propose that the *ideal* tensile-to-shear strength ratio ($I_{plasticity}^{slip}$) obtained from molecular dynamics is a reliable indicator of materials ability to resist fracture via slip-induced crack blunting: while the tensile strength should correlate with the stress required to open a crack, the shear strength should correlate with the stress necessary to nucleate *and* move a dislocation from a crack tip. We argue that the reliability of the $I_{plasticity}^{slip}$ descriptor lies in the fact that *ideal* strengths calculated via molecular dynamics explicitly include non-elastic and vibrational effects at a temperature of interest. Moreover, the computation of ideal tensile and shear strengths at finite temperatures is more practical than determination of surface and unstable stacking fault free-energies.

The $I_{plasticity}^{slip}$ descriptor accounts for the Schmid's factor ($m_{Schmid}$), which resolves the applied tension onto a slip plane inclined to the fracture plane. The natural starting point to test validity of the $I_{plasticity}^{slip}$ descriptor is to consider fracture on (001) surfaces, which have lowest formation energy in B1 TiN [39, 67] and TiN$_x$ [68]. Among the $\{110\}\langle 1\bar{1}0\rangle$, $\{001\}\langle 1\bar{1}0\rangle$, and $\{111\}\langle 1\bar{1}0\rangle$ slip systems, known to operate in B1 nitrides and carbides at room and/or high temperature [69-74], the $\{110\}\langle 1\bar{1}0\rangle$ is the most frequently reported at 300 K [47, 70] and, for $\langle 001 \rangle$ tension, has the highest Schmid's factor (0.5). Thus, the focus of present AIMD and CMD investigations is on the competition between unstable $\{001\}$ crack growth and crack blunting via nucleation and motion of $\{110\}\langle 1\bar{1}0\rangle$ dislocations. Accordingly, we base our analyses on a $I_{plasticity}^{slip}$ descriptor defined as

$$I_{plasticity}^{slip} = m_{Schmid}^{\{110\}\langle 1\bar{1}0\rangle} \frac{\sigma_T^{\langle 001 \rangle}}{\gamma_S^{\{110\}\langle 1\bar{1}0\rangle}}. \tag{4}$$

The *ideal* tensile strength $\sigma_T^{\langle 001 \rangle}$ is the maximum tensile stress withstood by a pristine crystal prior to $\{001\}$ cleavage, while the *ideal* $\gamma_S^{\{110\}\langle 1\bar{1}0\rangle}$ shear strength corresponds to the maximum shear stress before activation of $\{110\}\langle 1\bar{1}0\rangle$ slip. The descriptor in **Eq. 4** is equivalent to the *ductility parameter* demonstrated in Ref. [75] to be a reliable indicator of ductility in nanowires under tension.

The shear strength $\gamma_S$ of dislocation-free single-crystal B1 TiN and TiN$_x$ is evaluated via simple-shear simulations along the $\{110\}\langle 1\bar{1}0\rangle$ slip system. It is plausible to expect that $\gamma_S$ values encompass both the stress required to nucleate *and* move a full dislocation. For the case of stoichiometric TiN at room temperature, the *ideal* shear strength value calculated by AIMD ($\gamma_S^{\{110\}\langle 1\bar{1}0\rangle}$ = 46.2 GPa) is much larger than the CMD one ($\gamma_S^{\{110\}\langle 1\bar{1}0\rangle}$ = 31.8 GPa), see **Table III**. However, an increase in temperature progressively brings AIMD and CMD values to closer agreement (**Table III**). Independent DFT investigations [14] of TiN shear-stress/strain relationships at 0 K return an ideal shear strength (≈31 GPa) and a yield point (≈14%) that match our CMD results at 300 K (**Table III**). Because nudging was employed to locate shear-instabilities during deformation, the DFT result of Ref. [14] is a solid support to our CMD-predicted shear strength $\gamma_S^{\{110\}\langle1\bar{1}0\rangle}$. The difference between the AIMD prediction of $\gamma_S^{\{110\}\langle1\bar{1}0\rangle}$ at 300 K and the value in Ref. [14] is attributed to different supercell sizes, k-point meshes, and cutoff energies.

The CMD-calculated *ideal* strengths of pristine TiN and TiN$_x$ crystal models exhibit relatively small differences (**Table III**). At room temperature, anion vacancies cause a reduction in both tensile (≈6%) and shear (≈10%) strength, which translates in a slightly larger $I_{plasticity}^{slip}$ value for TiN$_x$ (0.610±0.010) vs TiN (0.582±0.010), see **Table IV**. As described in **Sec. 3.2**, a slightly larger $I_{plasticity}^{slip}$ reflects a substantially improved resistance to fracture in TiN$_x$ due to more frequent nucleation and emission of $\{110\}\langle1\bar{1}0\rangle$ edge dislocations.

**3.1.C. Comparison of edge-dislocation properties in TiN and TiN$_x$.** The observation of nucleation and motion of dislocations during K$_I$-controlled simulations (**Fig. 2c,d**) and molecular dynamics (see below) motivates dedicated MS investigations of the properties of these defects. **Fig. 4c,d** shows relaxed $\{110\}\langle1\bar{1}0\rangle$ edge dislocation cores in TiN and TiN$_x$. The results are obtained for 3D-periodic supercells of constant thickness along the dislocation line (// to [001]) and distance between dislocation dipoles set to half the supercell [110]-size (≈20 nm).

The atomistic structure of the relaxed edge-dislocation core in TiN illustrated in **Fig. 4c** is consistent with that predicted by DFT for TiN [49] and alike the one observed by electron microscopy for B1 MgO ceramics (figure 2 in Ref. [76]). The core-structure in vacancy-ordered TiN$_x$ resembles that of TiN but with slightly broader atomic displacements along [$1\bar{1}0$] (**Fig. 4d**). The map of lattice distortions calculated for TiN is mirror-symmetric in the plane of view (**Fig. 4c**). In contrast, Nye analysis reveals a non-symmetric distribution of lattice distortions in TiN$_x$, which is due to a non-symmetric arrangement of vacancies around the core (**Fig. 4d**). For line-energy estimations, we increase the lateral supercell size from ≈5 nm up to ≈50 nm. **Fig. 4a** shows that, for [1$\bar{1}$0]-supercell-sizes >15 nm, the dislocation line-energies of B1 TiN and TiN$_x$ saturate to ≈5.6 and ≈5.0 eV/Å.

The supercells used for MS calculations of Peierls stress $\tau_P$ have a lateral size of ≈30 nm. The value $\tau_P^{TiN}$ = 1.27 GPa, **Fig. 4b**, is close to the range ($\tau_P^{TiN}$ = 1.3–1.4 GPa) obtained by DFT [49]. The agreement may be accidental due to the use of different supercell models. The results obtained for vacancy-ordered TiN$_x$ supercells ($\tau_P^{TiNx}$ = 1.74 GPa) indicate that greater shear stress is required to activate motion of {110}⟨1$\bar{1}$0⟩ edge dislocations in the nitrogen-deficient crystal, **Fig. 4b**. A higher $\tau_P$ is likely due to dislocation-pinning at vacancy sites. On the other hand, the fact that the line-energy estimated for TiN (5.6 eV/Å) is larger than that of TiN$_x$ (5.0 eV/Å), **Fig. 4a**, is indicative of lower probability for dislocation nucleation in the stoichiometric phase. Estimates of the properties of {110}⟨1$\bar{1}$0⟩ edge dislocations aid understanding of slip-induced plasticity (see **Sec. 3.2**).

**3.1.D. Mechanical properties of *small* notched models under tension.** The mechanical properties of notched TiN models (**Fig. S1**) are evaluated via AIMD simulations of mode-I tension (strain normal to the crack plane) at different temperatures. **Fig. 5** shows a comparison of AIMD and CMD snapshots of strained supercells at 300 and 1200 K. Results at 600 K and 900 K (not shown) are qualitatively similar. Irrespective of temperature, the simulations indicate that Ti-N bonds near the notch tips start breaking when the simulation box is strained by ≈20% (see red arrows in **Fig. 5**). Nevertheless, for temperatures of 300 K (and 600 K), TiN activates and sustains crack-bridging up to 40% strain (black arrows in **Fig. 5**). Crack-bridging is an extrinsic toughening mechanism (i.e., operates behind the crack) that has been experimentally shown for other brittle ceramics (see, e.g., electron microscopy micrographs for microindented (V,Si)N coatings; figure 5b1,c1 in Ref. [77]).

The agreement between CMD and AIMD results holds also for fracture strengths $\sigma_f$. Irrespective of temperature, $\sigma_f$ determined by AIMD are in the range ≈5.0±1.5 GPa (**Table S1**). Large error bars are due to limited NVT-sampling times (≈1.6 ps) at each strain. In CMD simulations, the estimated fracture strength of *small* notched TiN models are within AIMD uncertainty ranges (**Table S1**), which supports the reliability of the interatomic potential used in present CMD investigations.

### 3.2. Competition between brittleness and plasticity in flawed TiN and TiN$_x$

In **Sec. 3.1**, we have validated our CMD/MEAM description of TiN properties against AIMD and DFT results obtained for single crystals, edge dislocations, and *small* notched supercells. $K_{Ic}^G/K_{Ie}^R$ ratios in **Table I** indicate that TiN(001)[010] and TiN$_x$(001)[010] are brittle materials, which is confirmed $K_I$-controlled MS simulations of plates with atomically-sharp cracks (**Fig. 2a,b**). Nevertheless, TiN plate models with three-atom-thick cracks exhibit slip-induced plasticity (**Fig. 2c,d**), indicating that the crack sharpness has a strong influence on the mechanical behavior. In addition, a comparison of $K_{Ic}^G/K_{Ie}^R$, $K_{IIe}^{KC}$ (**Tables I and II**) and dislocation line-energies (**Fig. 4a**) indicate that the presence of anion vacancies facilitates dislocation nucleation and emission during mode-I and mode-II loading.

During static K-controlled loading, the outcome of the simulation is uniquely determined by the initial simulation setup. At finite temperatures, instead, different mechanisms may compete and thus affect the simulation outcome. This calls for statistical analyses of the material mechanical behavior at finite temperatures [78]. We aim at gaining a qualitative understanding of the system response by applying tensile deformation to notched supercells via CMD simulations (**Fig. 1**). The use of supercells of relatively small area $b \times h = 160$ nm$^2$ (**Fig. 1**) allows us to collect statistics at a feasible effort. This is not expected to qualitatively alter our conclusions, because the mechanical behavior of defective Ti-N during K-controlled loading remains unaltered for plate areas between ≈72 and 7300 nm$^2$. However, we verify that an increased supercell thickness (from 0.9 to 17.0 nm along the [010] direction in **Fig. 1**) does not significantly impact the probability to observe slip-induced plasticity. In this regard, experiments show that dislocation-mediated plasticity is pronounced in samples of 150 nm thickness, whereas nearly absent in 300-nm-thick samples [23].

In CMD tensile-testing simulations, the stress intensity at the notch tip is incremented by increasing the vertical length of the simulation box. The procedure avoids biasing the lattice vibrations, as it would occur in K-controlled simulations (**Sec. 2.1.C**). The relative speed of accumulation of tensile vs shear stress at the notch tip – which corresponds to different degrees of mode-I/mode-II stress intensities – is tuned by changing the width of the void region $V_{\vdash\dashv}$ in **Fig. 1**. The understanding of effects induced by mixed mode-I/mode-II loading on mechanical properties aids interpretation of the behavior of protective ceramic coatings during industrial operation.

Starting with supercells of 2.6 nm thickness, independent CMD tensile-tests are carried out for each $V_{\vdash\dashv}$ parameter at each investigated temperature. Tensile loading of notched TiN and TiN$_x$ with $V_{\vdash\dashv} > 0.4b$ leads, with few exceptions, to brittle fracture via either crack growth or sudden cleavage across the entire supercell. This is explained by the fact that $V_{\vdash\dashv} > 0.4b$ induce a *too rapid* accumulation of tensile stress in comparison to the rate of shear stress increase at the notch tip. Conversely, CMD testing of supercells with $V_{\vdash\dashv} \leq 0.4b$ exhibits competition between crack propagation and slip-induced plasticity. A separate analysis of atomic displacements in MS calculations reveals that elongation of supercells with $V_{\vdash\dashv} > 0.4b$ essentially corresponds to pure mode-I loading. In contrast, the displacements in tensile-strained supercells with $V_{\vdash\dashv} \leq 0.4b$ are compatible with mixed mode-I (≈85–95%) / mode-II (≈15–5%) loading. This suggests that accumulation of shear stress at the crack tip is responsible for the more frequent activation of slip-induced crack-blunting in supercells with $V_{\vdash\dashv} \leq 0.4b$.

An example of competition between brittleness and slip-induced plasticity is illustrated by **Figs. 6 and Fig. 7**. Independent (i.e., different initial atomic velocities) CMD tensile tests are performed for notched TiN with $V_{\vdash\dashv} = 0.4b$ at room temperature. Atoms are colored according to local values of uniaxial $\sigma_{zz}$ (**Figs. 6a and 7a**) and shear $\sigma_{xz}$ (**Figs. 6b and 7b**) stresses. The stress color-scales are logarithmic: small color variations correspond to a relatively large change in stress. Despite starting with supercells of equal shapes and using equal strain rates, the simulations evidence *qualitatively different* mechanical responses to stress. **Fig. 6** shows unstable crack growth, whereas **Fig. 7** illustrates a case of slip-induced crack blunting due to nucleation and emission of $\{110\}\langle1\bar{1}0\rangle$ edge dislocations from the notch tip. The phenomenon, known to operate in metals [64], is unusual for hard ceramics. Below we provide an interpretation for the contrasting mechanical behaviors of **Figs. 6 and 7**.

**Figs. 6a,b and 7a,b** show CMD snapshots of TiN supercells elongated by 5% at timeframes closely preceding (≈1 ps) crack growth (**Fig. 6**) or slip-induced plasticity (**Fig. 7**). A strain of 5% produces tensile- and shear-stresses which have local maxima at the notch tips. However, thermal fluctuations temporarily generate greater tensile-to-shear stress ratios around the notch tip of **Fig. 6a,b** than in **Fig. 7a,b** (compare shear stress color intensities in **Figs. 6b and 7b**). In the case presented in **Fig. 6a**, a high tensile stress causes breakage of Ti-N bonds (**Fig. 6c**), ultimately leading to brittle fracture (**Fig. 6d**). Conversely, an increase in tensile stress accompanied by a comparably rapid accumulation of shear stress (**Fig. 7a,b**) assists nucleation of an edge dislocation (**Fig. 7c**). A further increase in strain triggers dislocation emission, which blunts the notch and impedes crack growth (**Fig. 7d**). The beneficial effect of shear stress accumulation on slip-induced plasticity at crack tips is also indicated by K-controlled simulations. **Fig. 2** shows that, while TiN plates with atomically-sharp cracks fail by crack propagation, TiN plates with 3-atomic-layer-thick cracks nucleate and emit dislocations, which is similar to the results in **Fig. 7**. In **Fig. S5** of the **SM** [20], we analyze the atomic stresses in the plates of **Fig. 2**. Subject to an equal mode-I load ($K_I$ = 1.4 MPa √m), large TiN plates with cracks of different heights accumulate approximately the same tensile stress **Fig. S5a,a1,b,b1**. However, as indicated by colored arrows in **Fig. S5d,d1**, the tip of the 3-atomic-layer-thick crack is subject to greater shear stress, which is considered the reason for dislocation nucleation and emission observed in **Fig. 2c,d**.

**Fig. 8** describes a typical pathway for dislocation nucleation and emission from a notch tip. In **Fig. 8a,a`**, the supercell box is elongated by 8%. After ≈1 ps at 8% strain, the material dissipates part of accumulated stress via rupture of all Ti-N bonds that connect [010] lattice rows labelled as **A** and **B** in **Fig. 8b`**. Due to sufficiently large shear stress at the notch front, part of the crystal below the slip plane penetrates the lattice creating a stacking fault, see **Fig. 8c,c`**. The defect remains at the notch tip until the uniaxial strain is increased to 9%, **Fig. 8 d,d`**, which leads to formation of a $\{110\}\langle 1\bar{1}0\rangle$ edge dislocation. Subsequently, the dislocation core moves at an average speed of ≈22 Å/ps across the entire crystal (see **Fig. 8 e,e`,f**) and exits within ≈3.5 ps from the bottom-right surface. The notch height – two atomic layers in **Fig. 8a,a`** – increases to three atomic layers in **Fig. 8e,e`**.

Although not directly comparable to mechanical tests by Kumar et al. [23], CMD results of dislocation-driven crack blunting in TiN crystals at 300 K are consistent with the observations of that work. **Fig. 9a** shows the percentage of tensile tests exhibiting slip-induced crack blunting in notched TiN and TiN$_x$ vs the $I_{plasticity}^{slip}$ value computed at the same temperature. A total of 350 for TiN and 140 for TiN$_x$ CMD simulations is performed at each investigated temperature (300, 600, 900, and 1200 K). Remarkably, $I_{plasticity}^{slip}$ values display a nearly monotonic relationship with the likelihood that dislocation-induced plasticity is activated in TiN or TiN$_x$ at a given temperature (**Fig. 9a**). The results of **Fig. 9a** demonstrate that the *ideal* strength values calculated for single crystals can be used to forecast trends in mechanical response of defective lattice models. However, the effect of slip-induced plasticity may be of practical relevance only at the nanoscale. Indeed, the experiments of Ref. [23] showed remarkable dislocation activity only for samples with thickness below 300 nm. Additional CMD simulations at 300 K allow us to verify that slip-induced crack-blunting remains active in supercells with thickness up to 17.0 nm (≈ 250 000 atoms). We carried out 70 independent CMD tests (10 for each V$_{\vdash\dashv}$) for supercells with thicknesses of 0.9, 5.9, 8.5, and 17.0 nm. Results in **Fig. 9b** indicate that (except for the thinnest models) the probability for dislocation-mediated crack blunting remains constant with increasing supercell thickness.

Before conclusions, we briefly compare the mechanical response observed in TiN and TiN$_x$ notched models. The results in **Fig. 9a** indicate that the nitrogen-deficient phase is more resistant to fracture, due to improved plasticity. The effect is attributed to a lower energetic cost for nucleation and emission of dislocations (**Tables I, II, and Fig. 4a**). Once formed, dislocations are emitted relatively rapidly from the notch tip, because the locally accumulated shear stresses (**Fig. 7b**) are much larger than Peierls stresses (**Fig. 4b**). Dislocation nucleation and motion dissipate stress, thus preventing crack opening. We remind that the vacancy-ordered TiN$_x$ variant used in this work has not been previously observed but serves for supporting the analysis on the correlation between properties of pristine and defective-supercells.

## 4. Conclusions

Motivated by surprising experimental observation of dislocation-mediated flaw tolerance in TiN, we carried out *ab initio* and classical atomistic investigations to shed light on atomic-scale mechanisms that regulate the competition between brittle fracture and slip-induced crack-blunting in these ceramics. Molecular statics mode-I controlled simulations demonstrate the intrinsic brittleness of TiN with an atomically-sharp crack. However, the observation of dislocation nucleation and emission in TiN with a crack of 3 atomic-layer thickness suggests that the stress condition at the crack tip dictates the competition between brittle fracture and crack blunting. Moreover, calculations of dislocation line-energies as well as K$_{II}$-controlled molecular statics simulations indicate that the presence of anion vacancies facilitates dislocation nucleation and emission. Molecular dynamics simulations of notched supercell models under tension show that a *too rapid* accumulation of tensile stress at the notch front inevitably leads to crack growth and brittle failure. Conversely, a comparably rapid increase in shear and tensile stress may activate dislocation nucleation and emission, thus blunting the notch. Crack blunting, with associated stress dissipation, prevents (or delays) crack growth. We propose that a descriptor [$I_{plasticity}^{slip}$ in **Eq. 4**] readily obtainable from calculations of *ideal* strengths of pristine single-crystal phases may forecast trends in dislocation-plasticity and related improvement in resistance to brittle fracture of defective ceramics. The $I_{plasticity}^{slip}$ descriptor is suggested as convenient alternative to Griffith and Rice criteria for assessing the competition between brittleness and plasticity at finite temperatures.

## Acknowledgements

All simulations were carried out using the resources provided by the Swedish National Infrastructure for Computing (SNIC) – partially funded by the Swedish Research Council through Grant Agreement Nº VR-2018-05973 – on the Clusters located at the National Supercomputer Centre (NSC) in Linköping, the Center for High Performance Computing (PDC) in Stockholm, and at the High Performance Computing Center North (HPC2N) in Umeå, Sweden. Dr. Luis Casillas Trujillo at NSC is gratefully acknowledged for technical support. We gratefully acknowledge financial support from the Competence Center Functional Nanoscale Materials (FunMat-II) (Vinnova Grant No. 2022-03071), the Swedish Research Council (VR) through Grant Nº VR-2021-04426 and 2019–05600, the Swedish Government Strategic Research Area in Materials Science on Functional Materials at Linköping University (Faculty Grant SFO-Mat-LiU No. 2009-00971), and the Knut and Alice Wallenberg Foundation through Wallenberg Scholar project (Grant No. 2018.0194).

**Figures**

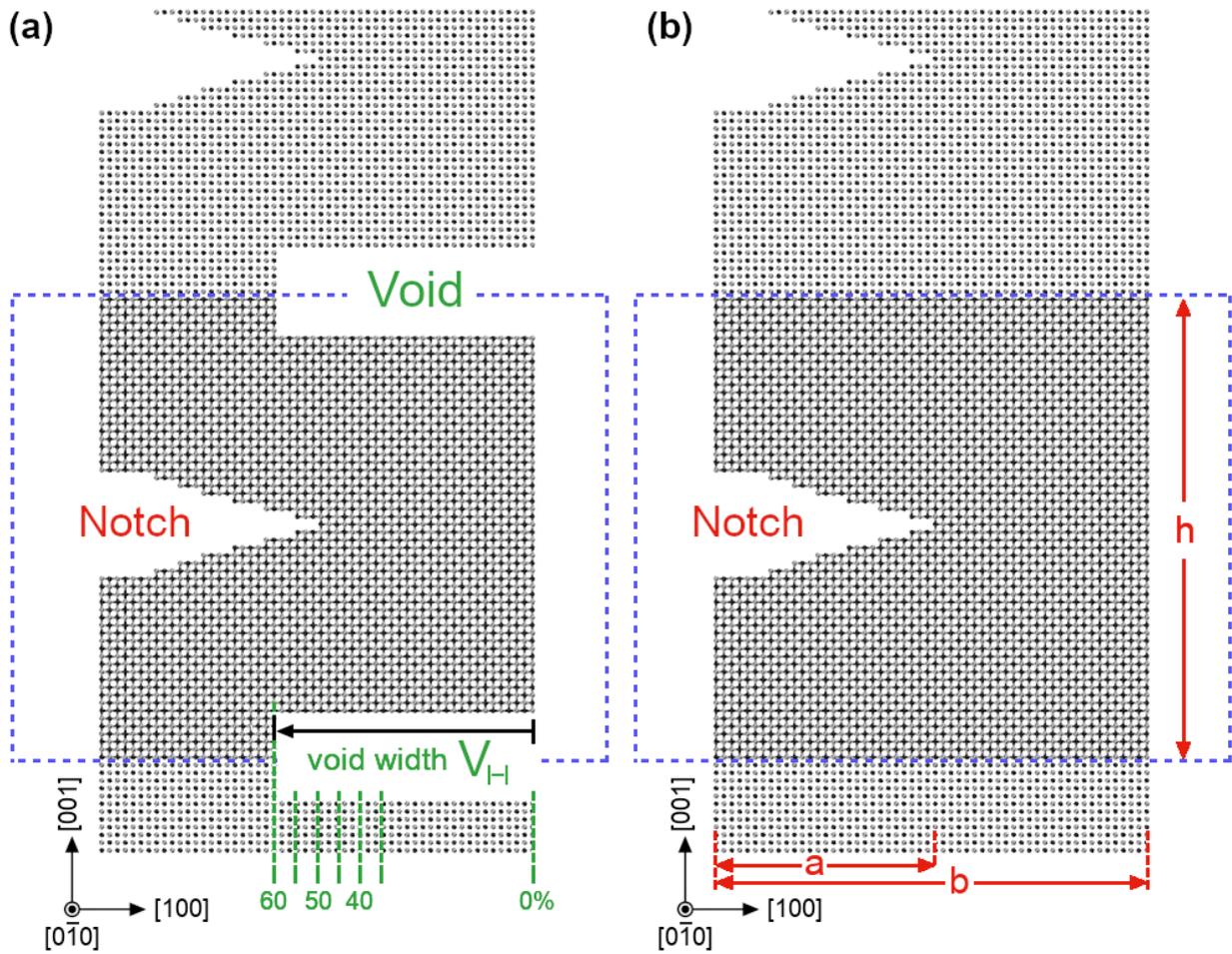

**Fig. 1**. Orthographic view of notched supercells used to investigate the mechanical behavior of TiN and TiN$_x$ during [001] tensile strain between 300 and 1200 K. The supercells are periodic along the tensile-strain direction [001] and crack-front direction [010]. The simulation-box boundaries are marked by dashed blue lines (vacuum separates supercell replicas along [100]). All supercells have equal notch geometry, shape $a/b \approx 0.5$ and $h = 12.8$ nm [see $a$, $b$, $h$ labels in **(b)**], but different void-widths $V_{|\text{-}|} = 0, 35, 40, 45, 50, 55, 60\%$ $b$ [see green dashed lines in **(a)**]. Panels **(a)** and **(b)** show $V_{|\text{-}|} = 0.6\ b$ and 0, respectively. The void width allows tuning the relative speed of tensile vs shear stress accumulation at the notch tip during tension, thus affecting the material's mechanical response.

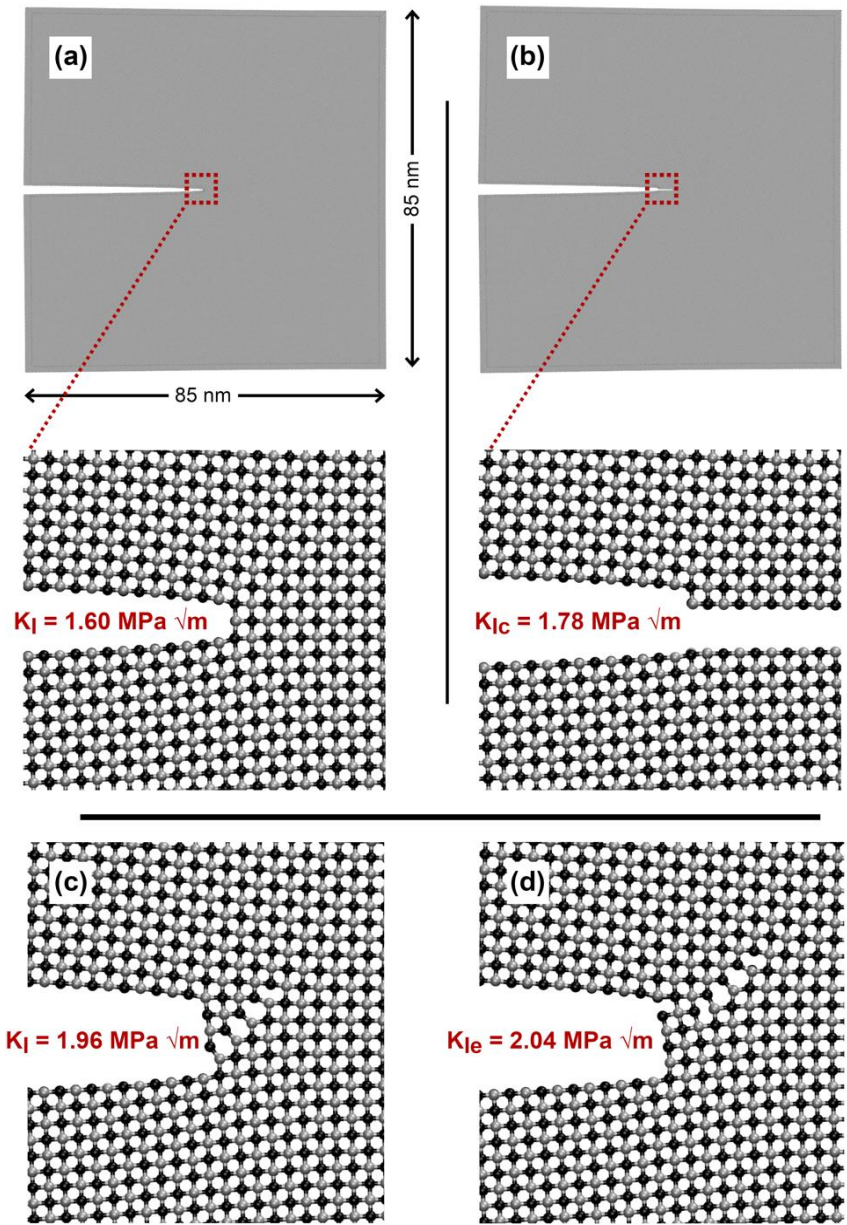

**Fig. 2**. $K_I$-controlled MS simulations of TiN(001)[010] plates with **(a)** and **(b)** atomically-sharp crack, **(c)** and **(d)** crack of 3-atomic-layer height. The plate areas are ≈7300 nm². **(b)** In the plate with atomically-sharp crack, crack growth occurs at $K_{Ic}^{KC}$ =1.78 MPa √m (**Video#1** in **SM** [20]). In the other case: **(c)** an edge dislocation nucleates at $K_I$ = 1.96 MPa √m and **(d)** moves toward the material interior for $K_{Ie}^{KC}$ = 2.04 MPa √m, thus increasing the crack thickness to 4 atomic layers (**Video#2** in **SM** [20]).

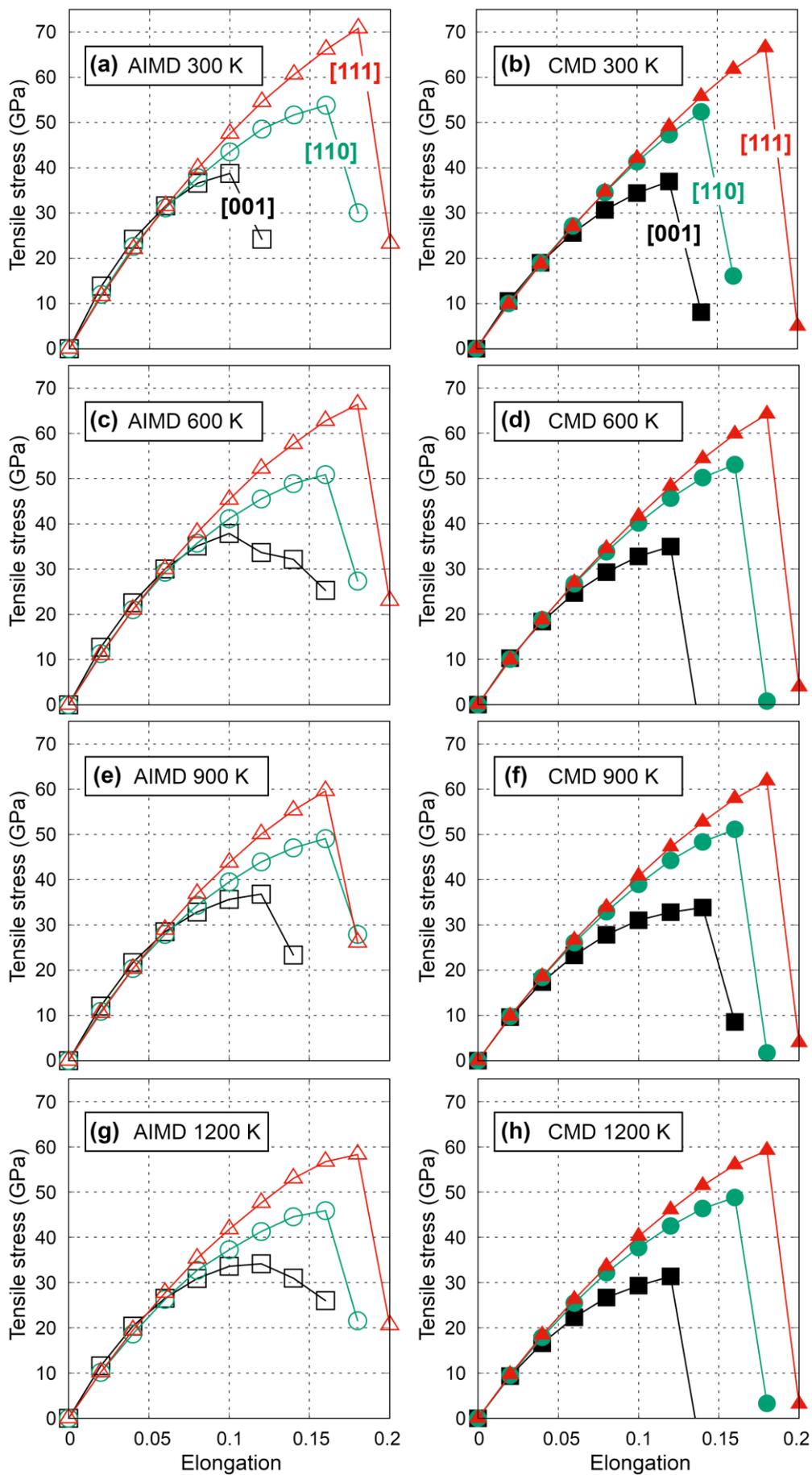

**Fig. 3**. AIMD **(a,c,e,g)** vs CMD **(b,d,f,h)** results for tensile deformation (Cauchy stress/strain) of stoichiometric defect-free TiN as a function of temperature. All curves end with the fracture point. AIMD stress strain data at room temperature presented in panel **(a)** are taken from Ref. [16].

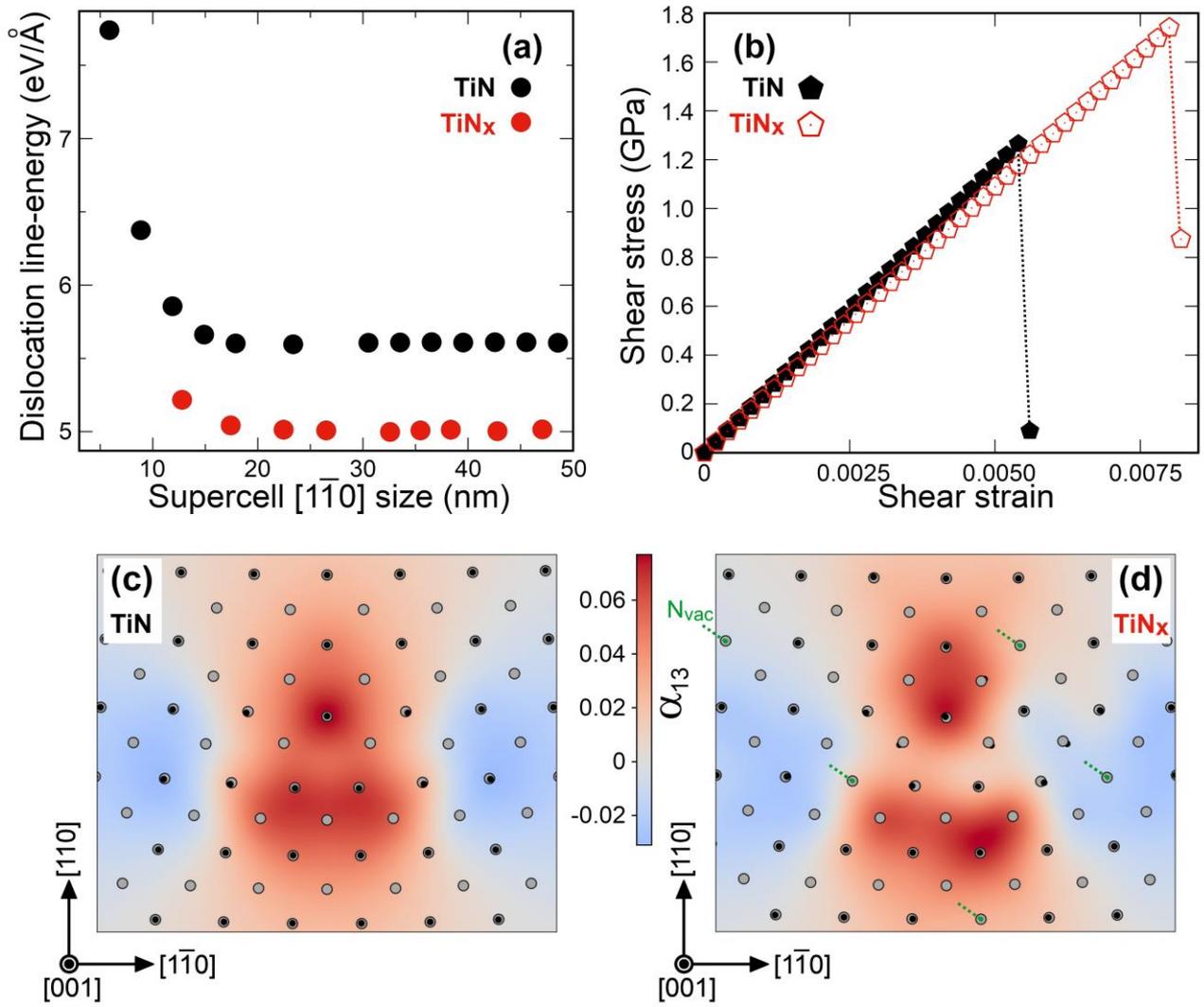

**Fig. 4**. MEAM results for {110}⟨1$\bar{1}$0⟩ edge-dislocations. Qualitative comparisons of line-energies **(a)** and Peierls stresses $\tau_P$ **(b)**. Relaxed core structures and contour plots of the distribution of $\alpha_{13}$ component of the Nye tensor in B1 TiN **(c)** and vacancy-ordered TiN$_x$ **(d)**. Silver and black spheres represent Ti and N atoms, respectively. Green dashed arrows in **(d)** indicate positions of nitrogen vacancies (N$_{vac}$) in the plane of view.

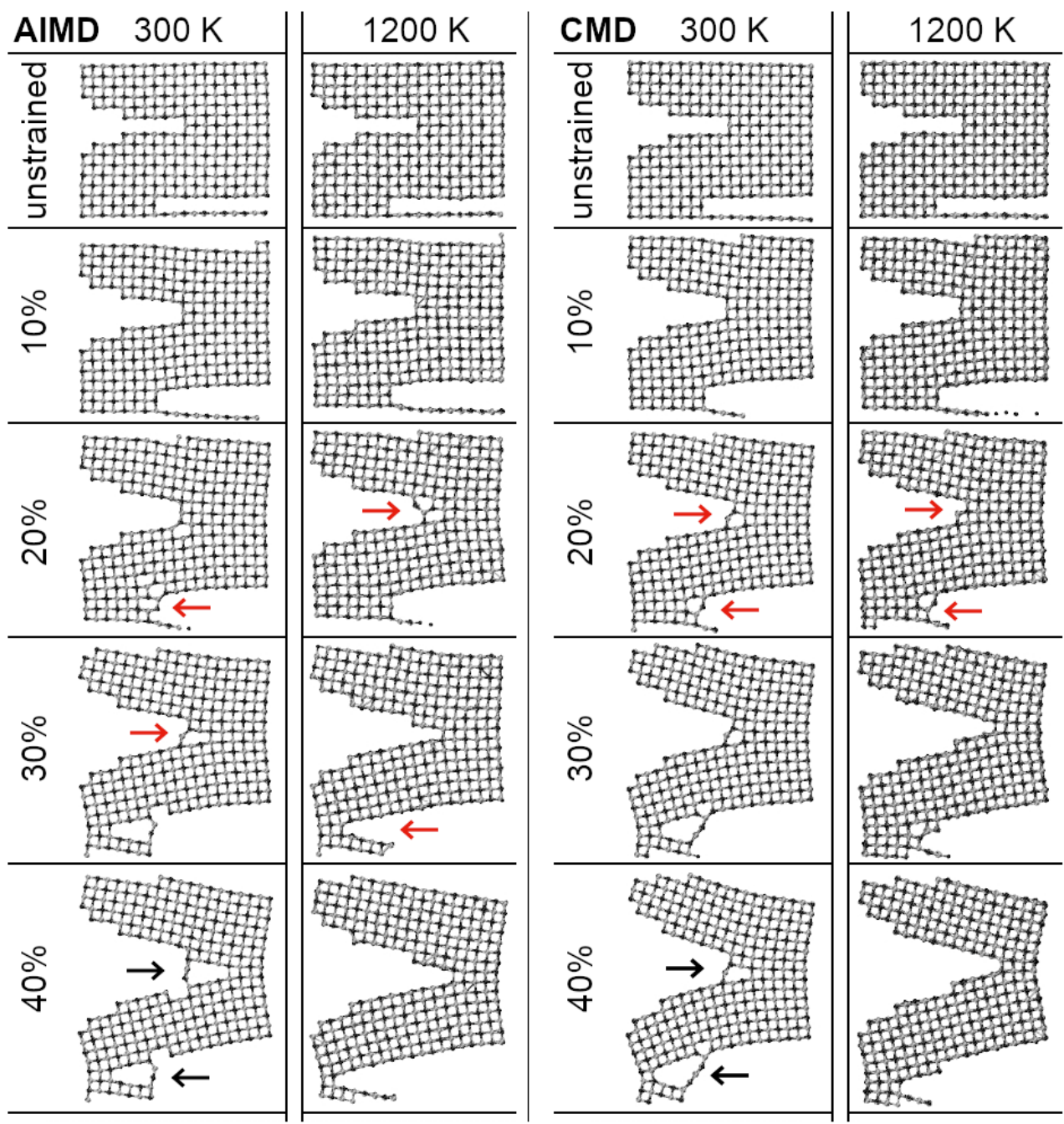

**Fig. 5**. AIMD and CMD snapshots for *small* notched TiN model (see **Fig. S1**) subject to tensile strain as a function of temperature. The percentages indicate vertical strain of the simulation box (which does *not* correspond to local deformation at crack tips). Black arrows indicate crack bridging. Red arrows indicate initial stages of Ti-N bond breakage at the notch tip. AIMD and CMD results at 600 K and 900 K (not shown) are similar to those presented in this figure. The chemical bonds have maximum lengths of 2.6 Å. Silver and black spheres represent Ti and N atoms.

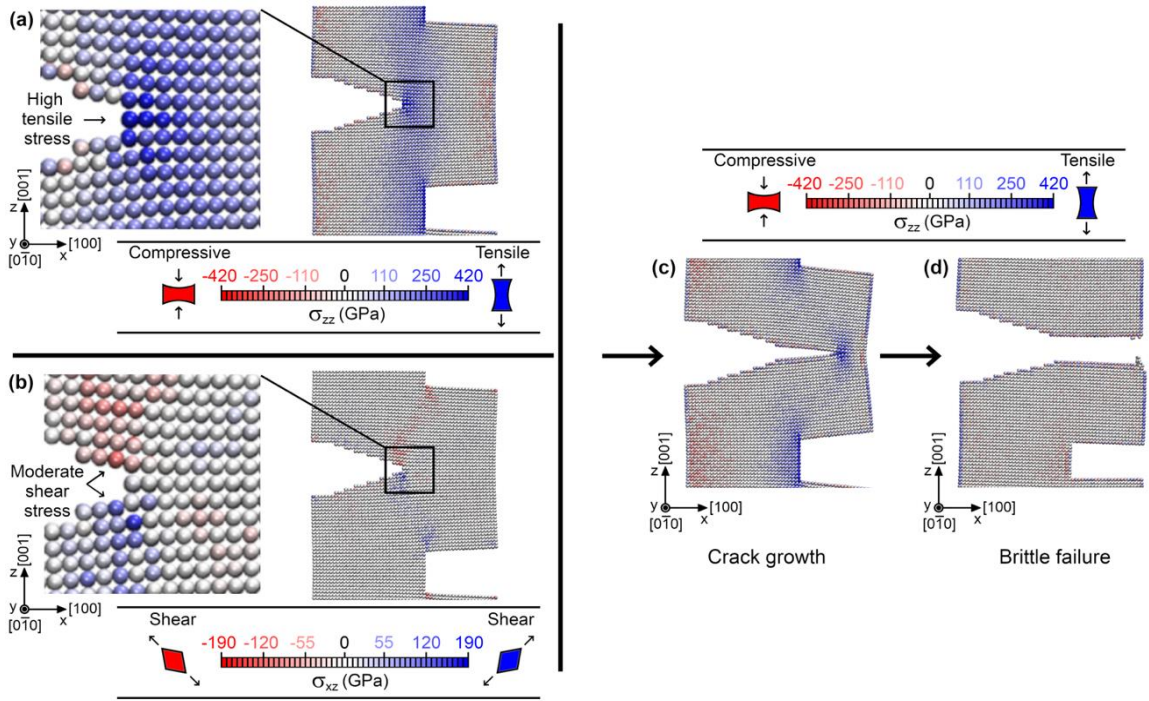

**Fig. 6**. Results of CMD tensile-strain simulations of a notched TiN supercell with void width $V_{\vdash\dashv} = 0.4b$ (see **Video#3** in **SM** [20]). Nitrogen and titanium species are not distinguishable. In **(a)** and **(b)**, the simulation box is strained by 5%. Panel **(a)** shows the distribution of uniaxial (compressive or tensile) stresses. Panel **(b)** shows local shear stresses. "High" tensile stress (blue color in **(a)**) combined with "moderate" shear stress (color scale in panel **(b)**) at the notch tip initiates crack growth **(c)** leading to brittle failure **(d)**. Note that only uniaxial-stress distributions are showed in panels **(c)** and **(d)**. The color-scales for uniaxial $\sigma_{zz}$ and shear $\sigma_{xz}$ stresses are logarithmic.

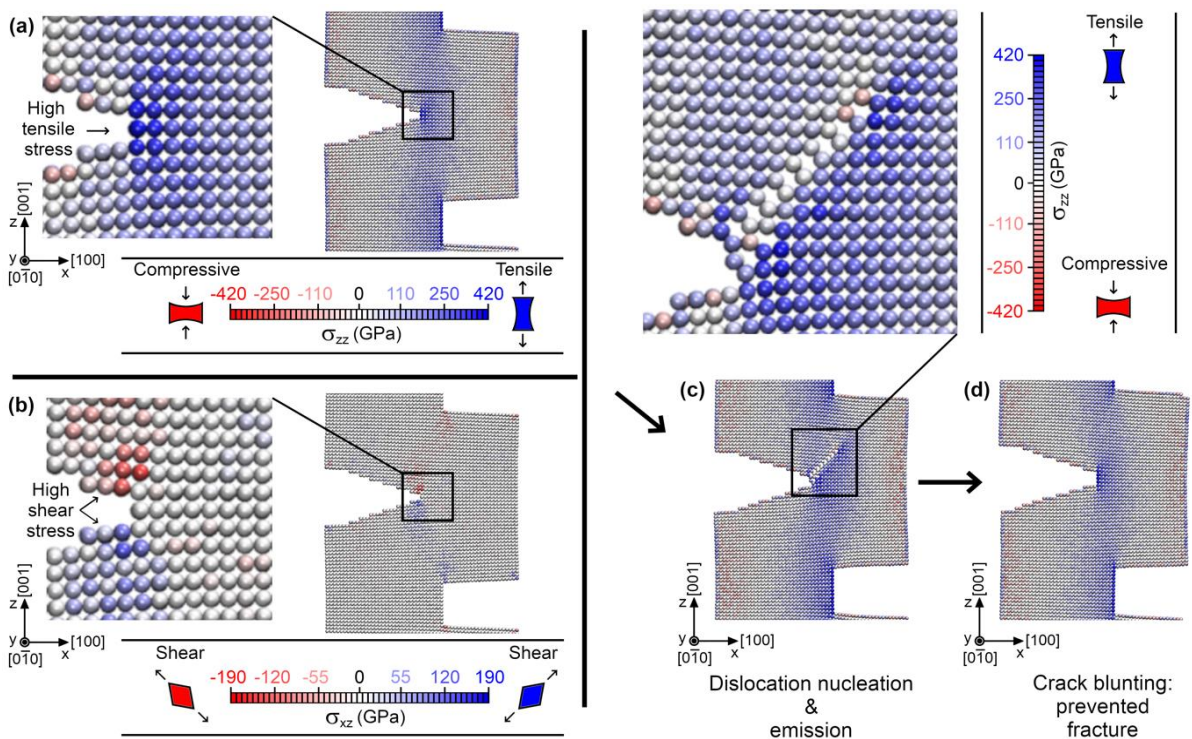

**Fig. 7**. Results of a simulation with different initial velocities than in **Fig. 6** (see **Video#4** in **SM** [20]). **(a)** High tensile stress combined with **(b)** relatively high shear stress assists **(c)** nucleation and emission of dislocations, which **(d)** blunts the crack and prevents fracture. In **(a)** and **(b)**, the simulation box is strained by 5%.

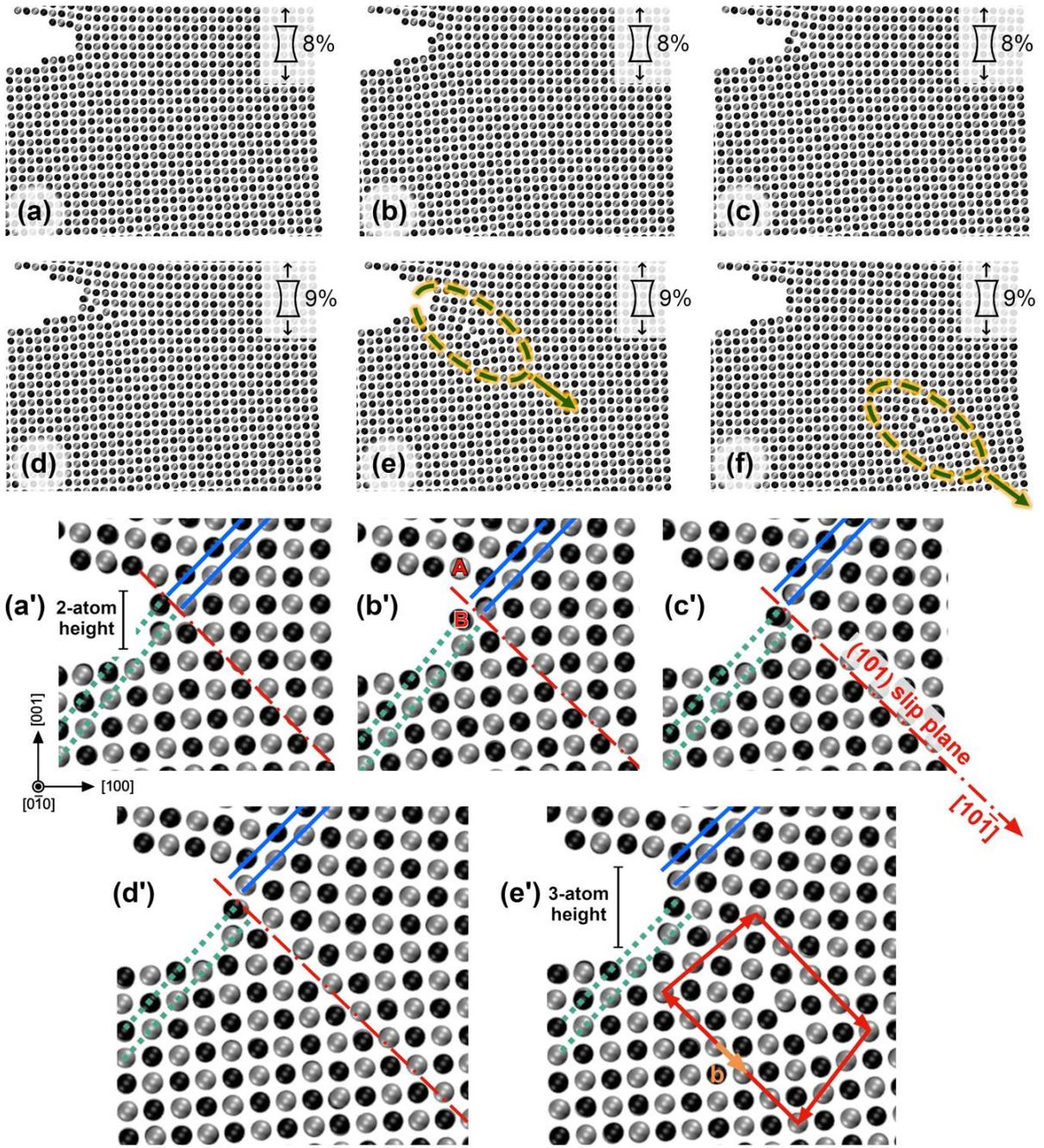

**Fig. 8**. Typical atomistic pathway for nucleation and emission of an edge-dislocation at notch tip of TiN subject to tensile strain **(a-f)**, with corresponding magnifications in panels **(a`-e`)**. Dash-dotted [ · — · — ] lines indicate a (101) slip plane. In **(a`)**: dotted green lines [ - - - - ] mirror solid blue lines [———] across the slip plane indicating aligned Ti and N [101] lattice rows prior to slip. The [001] strain of TiN supercells is 8% in **(a,a`,b,b`,c,c`)** and 9% in **(d,d`,e,e`,f)**. Snapshots **(a,a`)** show a two-atom-height notch tip, that is, one dislocation was previously emitted from the one-atom-sharp tip. The plots in **(b,b`)** evidence concerted breakage of Ti-N bonds linking **A**-**B** [010] lattice rows at the notch front. In **(c,c`)**, a portion of the crystal under the slip plane enters the lattice and creates a stacking fault defect: note that [ - - - - ] and [———] are no longer aligned. An increase in strain to 9% **(d,d`)** assists penetration of the stacking fault into the lattice via (101)[10$\bar{1}$] slip. The process generates a {110}⟨1$\bar{1}$0⟩ edge dislocation. The region of the crystal under the slip plane contains two extra (10$\bar{1}$) atomic layers. This can be understood by seeing **(d,d`)** that two lattice layers marked by [- - - -] lines overstepped the two planes indicated with blue lines [———] via displacement along the Burgers vector **b** = [10$\bar{1}$]. In **(e,e`)**, the dislocation core has moved toward the crystal interior. The Burgers circuit is shown in **(e`)**. Snapshot **(f)** shows that the dislocation core migrates rapidly under stress.

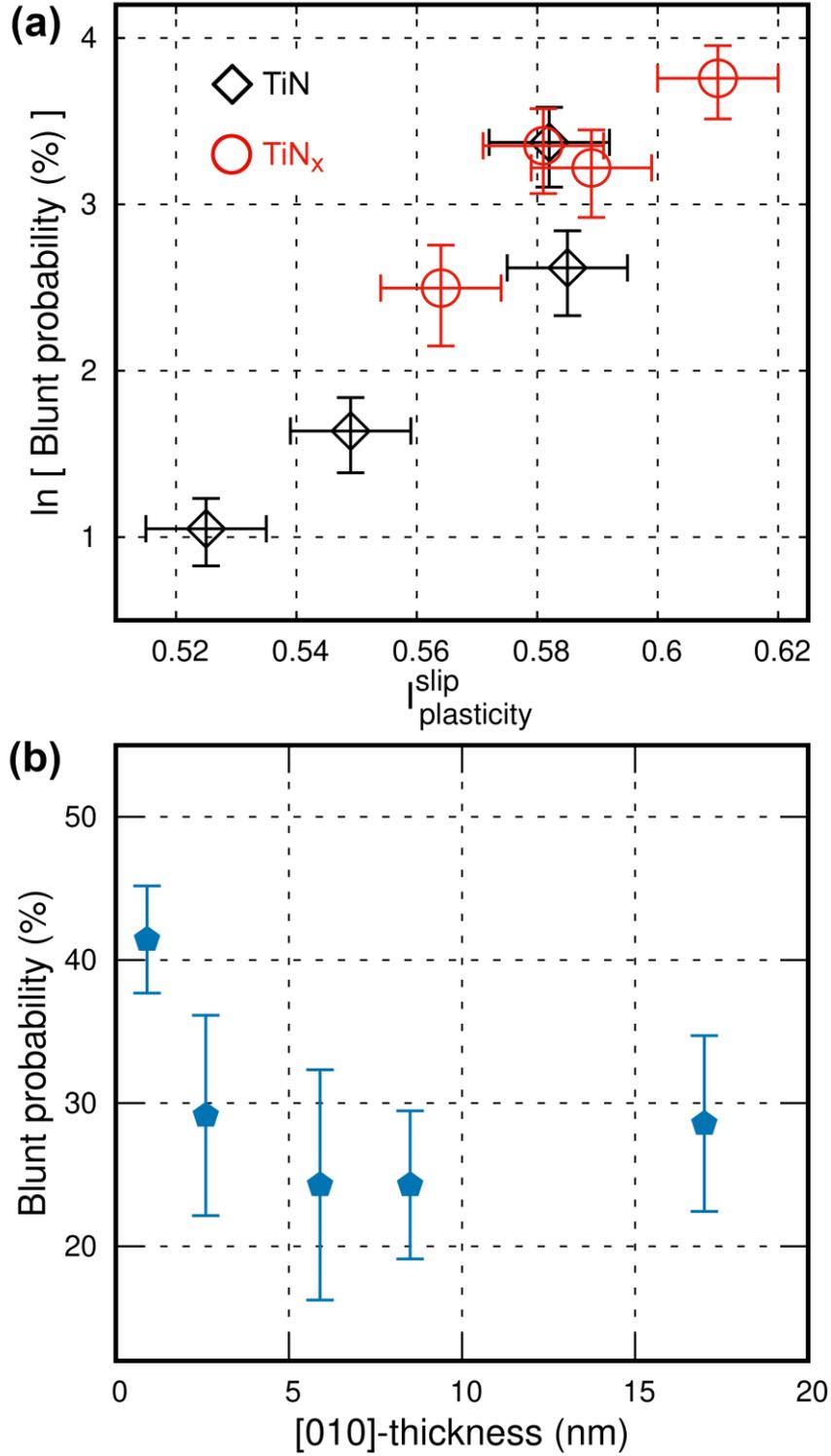

**Fig. 9.** (a) $I^{slip}_{plasticity}$ descriptor of probability of occurrence of dislocation-induced crack-blunting at the notch tip of initially dislocation-free TiN and TiN$_x$ lattices (model structures shown in **Fig. 1**). Each data point in **(a)** is obtained by CMD tensile tests at temperatures of 300, 600, 900, and 1200 K and for supercell thicknesses of 2.6 nm along the crack-front direction [010]. A log-scale representation is chosen to facilitate visualization of the trend. **(b)** Probability of occurrence of crack-blunting in TiN notched models at room temperature plotted as a function of the supercell thicknesses. 350 independent CMD runs (50 for each void width) are carried out for thickness of 2.6 nm. 70 simulations (10 for each void width) are carried out for thicknesses of 0.9, 5.9, 8.5, and 17.0 nm. The error bars are due to simulations in which the outcome is a mixture of plastic and brittle behavior.

**Tables**

|  |  | TiN |  |  |  | TiN$_x$ |
|---|---|---|---|---|---|---|
|  |  | DFT (present work) |  | DFT & experiments from literature | MEAM | MEAM |
|  |  | PBE | LDA |  |  |  |
| Elastic constants (GPa) |  |  |  |  |  |  |
| $C_{11}$ |  | 601 | 745 | 561 – 735 [a] | 613 [b] | 571 |
| $C_{12}$ |  | 124 | 143 | 93 – 165 [a] | 140 [b] | 142 |
| $C_{44}$ |  | 158 | 169 | 156 – 250 [a] | 165 [b] | 168 |
| Surface energies (J m$^{-2}$) |  |  |  |  |  |  |
| $E_{s,unrel}$ (001) |  | 1.54 | 2.00 | 1.53–1.76 [a] | 2.36 | 2.32 |
| $E_{s,rel}$ (001) |  | 1.23 | 1.70 | 1.06–1.30 [a] | 2.31 | 2.26 |
| $E_{s,unrel}$ (110) |  | 3.13 | 3.63 | 2.87–3.14 [a] | 3.39 | – |
| $E_{s,rel}$ (110) |  | 2.85 | 3.24 | 2.59–2.86 [a] | 3.26 | – |
| $E_{s,unrel}$ (111) |  | – | – | 4.58–5.62*; 5.08–5.45 [a] | 4.66 | – |
| $E_{s,rel}$ (111) |  | – | – | 3.30–4.37*; 4.59–4.95 [a] | 4.46 | – |
| Stacking fault energy |  |  |  |  |  |  |
| $\gamma_{usf}^{\{110\}\langle1\bar{1}0\rangle}$ (J m$^{-2}$) |  | 1.27 | 1.63 | 1.22 [c] | 1.89 | 1.77 |
| Mode-I | Cleavage |  |  |  |  |  |
|  | $\Lambda^{-1}_{22}$ (GPa) | 503 | 593 |  | 516 | 495 |
|  | **$K_{Ic\,(001)}^{G\,(001)[001]}$ (MPa m$^{½}$)** | **1.24** | **1.54** |  | **1.56** | **1.52** |
|  | Emission |  |  |  |  |  |
|  | $\Lambda^{\theta\,-1}_{11}$ (GPa) | 503 | 593 |  | 516 | 495 |
|  | $F_{12}(\theta)$ | 0.340 | 0.345 |  | 0.338 | 0.334 |
|  | **$K_{Ie\,(\bar{1}01)[101]}^{R\,(001)[010]}$ (MPa m$^{½}$)** | **2.35** | **2.85** |  | **2.92** | **2.80** |
|  | **$K_{Ic}^G/K_{Ie}^R$ (%)** | **52.8** | **54.0** |  | **53.4** | **54.3** |
| Mode-II | Emission |  |  |  |  |  |
|  | $\Lambda^{-1}_{11}$ (GPa) | 503 | 593 |  | 516 | 495 |
|  | **$K_{IIe\,(110)[1\bar{1}0]}^{R\,(110)[001]}$ (MPa m$^{½}$)** | **0.80** | **0.98** |  | **0.99** | **0.94** |

*DFT GGA and LDA surface energies of stoichiometric TiN(111) with dipole corrections [79].
a=Marlo et al. and refs therein [80]. DFT based on different exchange and correlation approximations.
b=Work of Ti-N MEAM parameterization [36].
c=[40] DFT with PBE approximation.
**Table I.** Properties of TiN and TiN$_x$ computed using Griffith and Rice criteria (**Eqs. 1,2,3**). Quantities of major interest are in bold.

|  | **TiN** | **TiN$_x$** |
|---|---|---|
| $K_{Ic\ (001)}^{(001)[001]}$ (MPa √m) | | |
| Griffith | 1.56 | 1.52 |
| K-controlled MS | | |
| Pure screening | Crack propagation 1.68 | – |
| Atomically-sharp | Crack propagation 1.78 | Crack propagation 1.82 |
| 2-atom thick | Nucleation 1.84 → fracture 2.20 | – |
| $K_{Ie\ (\bar{1}01)[101]}^{(001)[010]}$ (MPa √m) | | |
| Rice | 2.92 | 2.80 |
| K-controlled MS | | |
| 3-atom thick | Nucleation 1.96 → emission 2.04 | – |
| 5-atom thick | Nucleation and emission 2.04 | – |
| $K_{IIe\ (110)[1\bar{1}0]}^{(110)[001]}$ (MPa √m) | | |
| Rice | 0.99 | 0.94 |
| K-controlled MS | | |
| Atomically-sharp | Nucleation and emission 1.22 | Nucleation and emission 1.08 |
| $K_{Ic}$ (MPa √m) | | |
| Microcantilever bending *experiments | 1.2, 1.9, 2.5, 2.6 | – |

**Table II**. Comparison of Griffith and Rice criteria with $K_{Ic}$, $K_{Ie}$, and $K_{IIe}$ values calculated by $K_I$-controlled and $K_{II}$-controlled MS simulations for TiN and TiN$_x$. The MS results are obtained for plates with areas ≈7300 nm$^2$ (≈1.2 × 10$^6$ atoms). The $K_{Ic\ (001)}^{(001)[001]}$ determined by MS simulations (1.68 – 2.20 MPa √m) are within the range of fracture toughness values obtained by TiN microcantilever bending: 1.2 – 2.6 MPa √m *[81-84]. Note, however, that experimental $K_{Ic}$ results are affected by sample properties as, e.g., density and grain boundary structures.

| AIMD | 300 K | | 600 K | | 900 K | | 1200 K | |
| --- | --- | --- | --- | --- | --- | --- | --- | --- |
| TiN | $\delta_y$ (%) | $\gamma_S, \sigma_T$ (GPa) | $\delta_y$ (%) | $\gamma_S, \sigma_T$ (GPa) | $\delta_y$ (%) | $\gamma_S, \sigma_T$ (GPa) | $\delta_y$ (%) | $\gamma_S, \sigma_T$ (GPa) |
| **Shear** | | | | | | | | |
| $\{110\}\langle 1\bar{1}0\rangle$ | 16 ($\approx$14)[14] | 46.2 ($\approx$31)[14] | 14 | 39.3 | 14 | 37.0 | 14 | 34.9 |
| **Tensile** | | | | | | | | |
| $\langle 001\rangle$ | 10 | 38.8 | 10 | 37.9 | 12 | 36.8 | 12 | 34.2 |
| $\langle 110\rangle$ | 16 | 53.8 | 16 | 50.9 | 16 | 49.1 | 16 | 45.9 |
| $\langle 111\rangle$ | 18 | 70.8 | 18 | 66.4 | 16 | 59.6 | 18 | 58.3 |
| **CMD** | 300 K | | 600 K | | 900 K | | 1200 K | |
| TiN | $\delta_y$ (%) | $\gamma_S, \sigma_T$ (GPa) | $\delta_y$ (%) | $\gamma_S, \sigma_T$ (GPa) | $\delta_y$ (%) | $\gamma_S, \sigma_T$ (GPa) | $\delta_y$ (%) | $\gamma_S, \sigma_T$ (GPa) |
| **Shear** | | | | | | | | |
| $\{110\}\langle 1\bar{1}0\rangle$ | 14.0 ($\approx$14)[14] | 31.8 ($\approx$31)[14] | 14.0 | 29.8 | 15.3 | 30.4 | 16.0 | 29.7 |
| **Tensile** | | | | | | | | |
| $\langle 001\rangle$ | 12.0 | 37.0 | 12.0 | 34.9 | 12.7 | 33.3 | 12.0 | 31.2 |
| $\langle 110\rangle$ | 14.0 | 52.4 | 16.0 | 53.2 | 16.0 | 51.0 | 16.0 | 48.9 |
| $\langle 111\rangle$ | 18.0 | 66.5 | 18.0 | 64.1 | 18.0 | 61.8 | 17.3 | 58.1 |
| **CMD** | 300 K | | 600 K | | 900 K | | 1200 K | |
| TiN$_x$ | $\delta_y$ (%) | $\gamma_S, \sigma_T$ (GPa) | $\delta_y$ (%) | $\gamma_S, \sigma_T$ (GPa) | $\delta_y$ (%) | $\gamma_S, \sigma_T$ (GPa) | $\delta_y$ (%) | $\gamma_S, \sigma_T$ (GPa) |
| **Shear** | | | | | | | | |
| $\{110\}\langle 1\bar{1}0\rangle$ | 14.0 | 28.5 | 15.3 | 28.1 | 14.7 | 26.2 | 15.3 | 25.7 |
| **Tensile** | | | | | | | | |
| $\langle 001\rangle$ | 12.0 | 34.7 | 12.0 | 32.6 | 12.0 | 30.8 | 12.0 | 28.9 |
| $\langle 110\rangle$ | 16.0 | 49.0 | 16.0 | 47.2 | 16.0 | 45.2 | 15.3 | 42.6 |
| $\langle 111\rangle$ | 18.0 | 58.6 | 18.0 | 56.4 | 18.0 | 54.4 | 18.0 | 52.2 |

**Table III**. AIMD and CMD results for ideal tensile $\sigma_T$ and shear $\gamma_S$ strengths and yield points ($\delta_y$). For comparison, DFT results for TiN shear strength [14] (calculated using thick k-meshes and high cutoff energy) are reported in parentheses. All CMD-calculated values are averaged over 3 independent sets of runs and have statistical uncertainties ±1.0% (for $\delta_y$) and relative errors of ±5% for tensile and shear strengths. Uncertainties on AIMD values are ±2% (for $\delta_y$) and ±10% for tensile and shear strengths.

|          | $I^{slip}_{plasticity}$ | | | |
|----------|-------------|-------------|-------------|-------------|
|          | 300 K       | 600 K       | 900 K       | 1200 K      |
| AIMD TiN | 0.42±0.07   | 0.48±0.07   | 0.50±0.07   | 0.49±0.07   |
| CMD TiN  | 0.582±0.010 | 0.585±0.010 | 0.549±0.010 | 0.525±0.010 |
| CMD TiN$_x$ | 0.610±0.010 | 0.581±0.010 | 0.589±0.010 | 0.564±0.010 |

**Table IV**. AIMD and CMD-calculated $I^{slip}_{plasticity}$ values (see **Eq. 4** and **Table III**). For CMD, these values are averages of 3 independent sets of runs of ⟨001⟩–tensile and {110}⟨1$\bar{1}$0⟩—shear deformation of single-crystals. Confidence ranges are statistical uncertainties on calculated *ideal* strengths.